# Reemergence of Trampolining in a Leidenfrost Droplet

Pranjal Agrawal[a], Gaurav Tomar[b], Susmita Dash[b]

[a]Interdisciplinary Center for Energy Research, Indian Institute of Science, Bangalore

[b]Department of Mechanical Engineering, Indian Institute of Science, Bangalore

*Abstract*

The levitating Leidenfrost (LF) state of a droplet on a heated substrate is often accompanied by fascinating behaviors such as star-shaped deformations, self-propulsion, bouncing, and trampolining. These behaviors arise due to the vapor flow instabilities at the liquid-vapor interface beneath the droplet at sizes typically comparable to the capillary length scale of the liquid. Here, we report on the spontaneous bouncing, trampolining, and hovering behavior of an unconstrained LF water droplet. We observe that a droplet exhibits intermittent increase in bouncing height at specific radii and subsequent reduction in the height of bounce leading to a quiescent LF state. The reemergence of the trampolining behavior from the quiescent hovering state without any external forcing is observed at sizes as low as 0.1 times the capillary length. We attribute the droplet bouncing behavior to the dynamics of vapor flow beneath the LF droplet. We propose that the trampolining behavior of the droplet at specific radii is triggered by subharmonic and harmonic excitation of the liquid-vapor interface. We attribute the intermittent trampolining events to the change in the natural frequency of the droplet and the vapor layer due to evaporative mass loss. This proposed mechanism of resonance-driven trampolining of LF droplets is observed to be applicable for different liquids irrespective of the initial volume and substrate temperatures, thus indicating a universality of the behavior.

**Introduction**

The formation of a thin vapor film beneath a droplet before it comes into contact with a superheated substrate is known as the Leidenfrost (LF) effect and was reported first in 1756 by JG Leidenfrost[1]. The Leidenfrost state of a droplet on a substrate is marked by a significant increase in the total evaporation time [2–4]. It is accompanied by several interesting dynamics ranging from strong internal convection, directional transport, and droplet bouncing[5–7]. While the Leidenfrost state of the droplet is detrimental to heat transfer applications, the droplet mobility due to the absence of contact line has potential in applications related to pharmaceuticals[8,9], power engineering[10,11], microreactors, and the formation of supra particles[9,12]. The internal convection inside LF droplets transitions from an axisymmetric toroidal flow to single roll motion resembling solid body rotation as the droplet becomes smaller[5,13]. The high internal flow in LF droplets have been explored for applications in microscale chemical reactors[8]. In addition, the single roll internal motion was reported to cause self-propulsion of LF droplets on substrates and was demonstrated to be capable of cleaning contaminants on a heated substrate[5,14,15].

In 1966, Watchters et al.[16] noted that although the droplet looks calm to the naked eye, their images showed a blurred liquid-air interface, suggesting that the droplet was in motion (probably bouncing). The bouncing was attributed to liquid-solid contacts in the film boiling regime[17]. While droplets larger than the capillary length scale levitate and stay stationary on the substrate, smaller droplets are shown to demonstrate bouncing[18]. For low impact speeds ($v$), associated with dispensing of a liquid droplet on a substrate, the Weber number, $We(= \rho_l v^2 R/\gamma)$ is less than 1; here $\rho_l$ is the density of liquid, $R$ is the droplet radius, and $\gamma$ is the interfacial tension of the liquid-vapor interface. Quasi-elastic rebound, with coefficient of restitution almost equal to 1, of LF droplets at such low Weber number suggests negligible dissipations when the droplet is in contact with the substrate [19–21].

A gently placed droplet on a superheated substrate ($We \ll 1$) has recently been shown to exhibit bouncing and trampolining for Bond number, $Bo = (R/l_c)^2$ ranging from $0.04 - 0.5$, where $l_c \left(= \sqrt{\frac{\gamma}{\rho_l g}}\right)$ is the capillary length of the liquid[7,22]. The transition from bobbing (sideways droplet motion) to bouncing in a levitating water droplet was reported to occur when the Froude's number, $Fr(= \frac{v_{int}}{\sqrt{2gR}}) \sim 1$ where $v_{int}$ is the internal flow velocity and $R \sim 1.8$ mm suggesting that bouncing occurs when the inertia of the internal liquid flows overcomes gravity[22]. It was also suggested that the periodic instances of increase in the amplitude of droplet oscillations as the droplet evaporates from $R = 1.75$ mm to $R = 0.675$ mm occurs due to resonance between the frequency of droplet internal flow and the natural frequency of the droplet. The trampolining behavior of water droplets was attributed to the overpressure in the vapor pockets resulting due to fluctuations in vapor layer thickness[7]. These deformations were reported to be a function of the modified capillary number ($Ca^*$ scales as $R^{2.75}$), suggesting a balance between the shear stress and surface tension forces acting at the liquid-vapor interface of the LF droplet[7]. As the droplet size reduces such that the $Bo < 0.04$ and $Ca^* < 0.1$, the droplet was observed to behave as a hard sphere, and the bouncing ceased due to the absence of interface deformation caused by the draining vapor[7]. However, we observe that although the trampolining decays down as the LF droplet becomes small ($Bo = 0.02$, and $Ca^* < 0.1$) and comes to a hovering state, the bouncing reemerges and the droplet trampolines again. While there are a few reports on the non-monotonic spontaneous bouncing, trampolining, and sudden reduction in the bouncing behavior of LF droplets[7,22], the cause of the sudden increases in the amplitude and the reemergence of trampolining in small droplets, even at $Bo < 0.04$ is still not clear.

Here, we report on the bouncing and trampolining including the periodic instances of a sudden increase in the droplet oscillation amplitude during evaporation of an unconstrained LF water droplet on a superheated substrate till it reduces to very small droplet radius ($Bo \sim 0.02$; corresponding to droplet radius $R \sim 0.3$ mm for water). Interestingly, in the cases of water and water-glycerol droplets with initial volume 10 $\mu$L, the bouncing behavior ceases at smaller sizes of droplet and re-initiates spontaneously. We propose that the reinitiation of the increase in the oscillation amplitude of LF droplets occurs due to resonance when the natural frequency of the vapor layer is in multiples of half of the droplet's natural frequency (Rayleigh frequency). The presence of resonating frequencies also explains the experimental observations related to trampolining and bouncing of the droplets at specific radii and the reemergence of trampolining at small droplet diameters. It also suggests that continuous trampolining of LF droplets is restricted by evaporation-led reduction in droplet volume that in turn changes the droplet's natural frequency. A parametric investigation using different substrates and liquids, at varying initial droplet volume and substrate temperature indicates that the phenomenon of resonance-driven bouncing dynamics in LF droplets is universal.

**Observations and Discussion**

We place a water droplet with an initial volume ~ 6 - 30 $\mu L$ (radius ~ 1-1.9 mm) on a heated concave quartz substrate maintained above the LF temperature. For small droplet size with diameter < $l_c$, we observe vertical oscillations or bouncing behavior of the droplet till it completely evaporates or lifts off from the substrate[7,22,23]. During the LF bouncing, several trampolining instances are observed where the amplitude of bouncing suddenly increases and the height of bouncing amplifies with each subsequent jump (Fig 1(a), inset, SI Fig S1). The number of times ($n$) the droplet demonstrates such behavior during the entire period of evaporation is dependent on the initial volume of the droplet (for initial size < $l_c$). For instance, $n$ = 2 -4 for initial droplet volumes ($V_i$) of 6 - 12 $\mu$L; $n$ remains invariant at 4 for $V_i$ = 22 and 30 $\mu$L (SI Fig S2). The instances of the occurrence of trampolining phenomenon when the bouncing height suddenly increases followed by reduction in the bouncing height continues till the droplet reaches a hovering state which is characterized by almost negligible bouncing for droplets of water and water-glycerol mixture (gly10, gly25 and gly50) (Fig 1a). The reduction in bouncing height occurs due to viscous dissipation in the droplet as it contacts the substrate which are also influenced by the shape of the droplet (prolate or oblate) during impact[19]. The hovering state was not observed in ethanol, acetone, and Novec 7000 owing to lower interfacial tension of the liquid which decreases the resistance to deformations of the droplet (Fig 1(b, c)). The hovering diameter is a function of the initial volume, and goes to as small as $R \sim 0.175$ mm ($Bo = 0.01$) for a water droplet of initial volume 6 $\mu L$. For this size of the LF drop, $Ca^* \cong 0.008$, is significantly smaller than $Ca^* \sim 0.1$, for which the vapor shear-induced ripples at the liquid-vapor interface of the droplet were reported to exist[7]. The reemergence of the trampolining phenomenon at such a smaller size suggests a mechanism different than that related to the vapor shear.

To investigate the mechanism of the reemergence of trampolining phenomenon, we performed particle image velocimetry to observe the flow behavior inside the droplets along with the interferometric measurements of the vapor gap in the case of LF droplets of water and water – glycerol mixtures. It is known that the droplets in a LF state or on a heated superhydrophobic substrate show different modes in the internal motion depending on the droplet size and liquid properties[13,24]. For LF water droplets, at higher sizes (diameter > $l_c$), a toroidal vortex flow is observed which changes to a single roll as the droplet evaporates (diameter ≤ $l_c$)[13]. Visualization of the internal flow in an unconstrained LF droplet, such as that used in our experiments, is challenging due to the bouncing of the droplet. However, the relatively quiescent state of the LF droplet in the hovering regime enabled flow visualization along with vapor layer characterization (SI Fig S3). In the hovering state, we observe a single convection roll in a water droplet, which continues as the droplet evaporates and shows the reemergence of trampolining. The gly25 droplets had negligible internal flow but still exhibited the reemergence of droplet trampolining, which indicates that the droplet internal flow does not explain the phenomenon of trampolining (SI Fig S3). We also observe the vapor layer profile to be devoid of any ripples, indicating that the shear stress is not the only driving mechanism for trampolining (SI Fig S4, S5).

We observe that for an evaporating LF droplet of a given liquid, the instances of trampolining occurred at the same average instantaneous radii, irrespective of the initial volume and substrate temperature (SI Fig S6). For instance, in a water droplet an increase in bouncing amplitude can be observed at a radius ~ 1.3 ± 0.1 mm irrespective of the temperature of the substrate and its initial radius (SI Fig S6a). As the droplet evaporates, the size continuously reduces, and trampolining occurs at radii of 0.89 ± 0.12 mm, 0.52 ± 0.15 mm, and 0.32 ± 0.12 mm (Fig 1a, SI Fig S6a). The image analysis of the droplets of various liquids with different initial volumes on a quartz substrate maintained above the LF temperature shows that trampolining occurs at discrete radii when $R < l_c/\sqrt{2}$ (Fig 2).

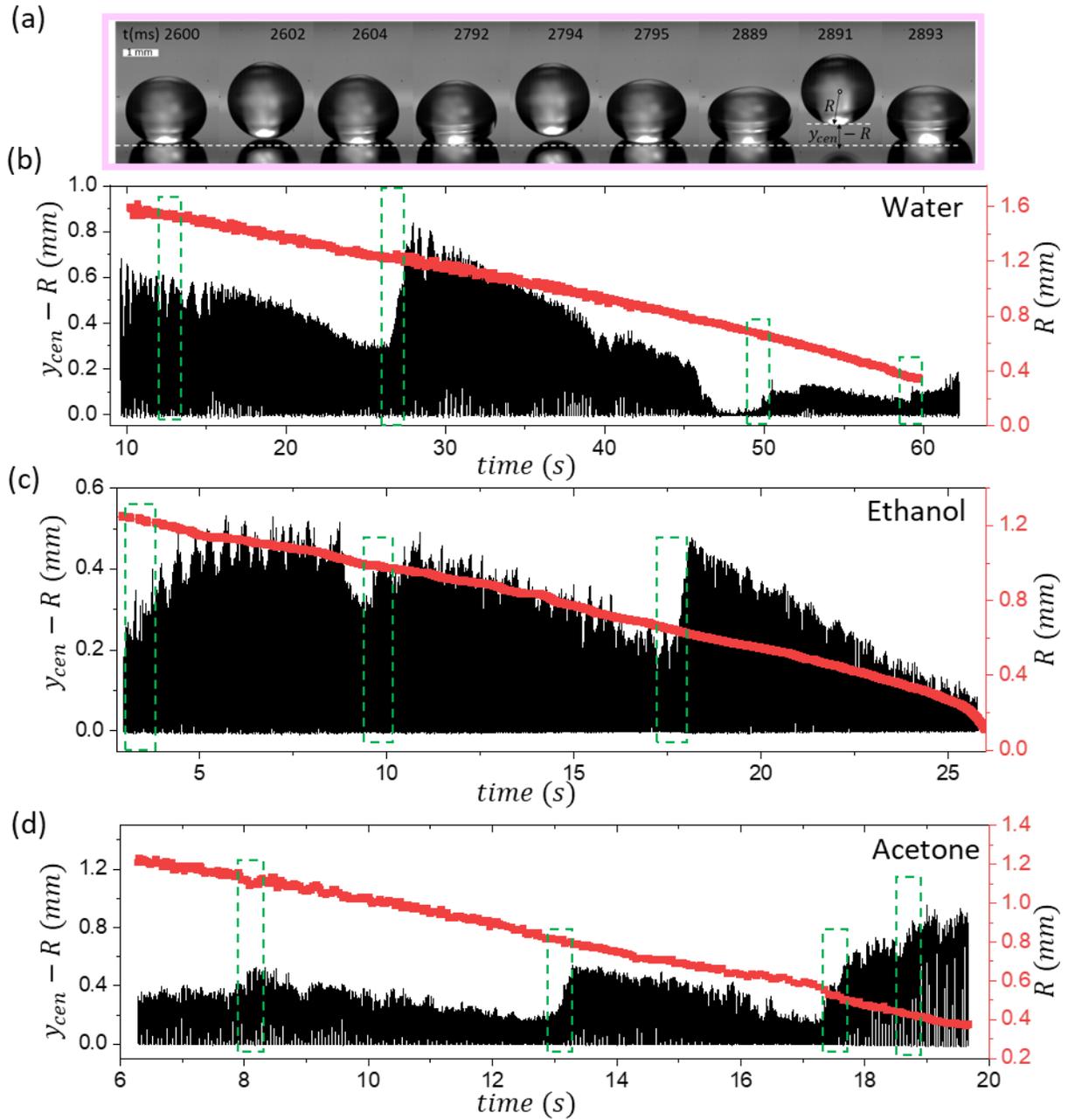

*Figure 1: (a) An instance of a water droplet trampolining on a heated substrate at $T_{sub}$ = 360 °C. Multiple instances showing an increase in the amplitude of bouncing of droplets of (b) water ($V_i$ – 15 μL) at*

*substrate temperature T$_{sub}$ = 360 °C; (c) ethanol (V$_i$ – 9 µL) at T$_{sub}$ = 290 °C, (d) acetone (V$_i$ – 9 µL) at T$_{sub}$ = 240 °C. The green dashed regions indicate the average radii at which trampolining is observed.*

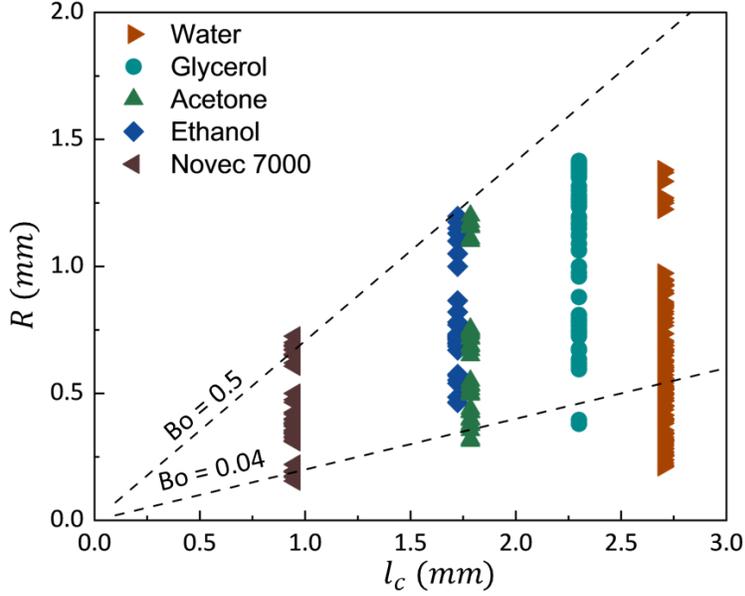

*Figure 2: Radii of the droplets at which trampolining is observed for different liquids with different initial volumes on a quartz substrate maintained at different temperatures.*

Figure 3a shows the bouncing dynamics of a water droplet ($V_i = 6\ \mu L$) in three different regimes. Regime 1 corresponds to the decaying oscillations where the droplet bouncing amplitude reduces and it eventually goes to a hovering state at R ~ 0.6 mm. In regime 2, the droplet stays in the hovering state as it evaporates from R ~ 0.6 mm to R ~ 0.51 mm. Regime 3 is marked by increased bouncing height and trampolining as the droplet size reduces beyond R = 0.51 mm. The frequency of bouncing of the LF water droplet in regime 1 and regime 3 is calculated using Fast Fourier transform (FFT) analysis in MATLAB to determine the dominant frequencies of bouncing (Fig 3a). In regime 1, dominant frequencies are observed at $f = 98\ Hz$ and 190 Hz (Fig 3b) while in the trampolining regime 3, peaks are at $f$ = 117 Hz and 235 Hz (Fig 3c). In the hovering state of the droplet, as the radius $R$ of the droplet reduces from ~ 0.6 mm to 0.51 mm, the Rayleigh frequency for quadrupolar droplet oscillation $f_{Rayleigh} = (1/2.65)\sqrt{\gamma/\rho_l R^3}$ ranges from 200 to 254 Hz.[19] Interestingly, in regime 1 and regime 3, that occurs before and after the hovering state, the dominating frequency of droplet bouncing are observed to be approximately half of $f_{Rayleigh}$.

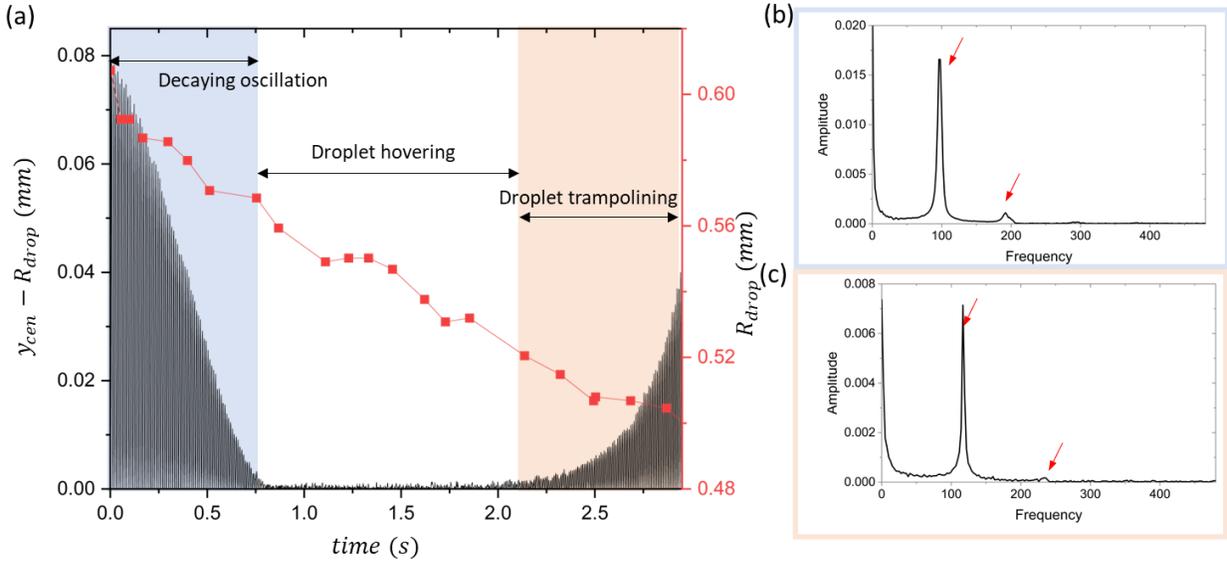

*Figure 3: (a) The temporal variation in droplet radius (red) and bouncing height (in black) of a LF water droplet of initial radius = 1.4 mm on a substrate at $T_s$ = 370 °C, t = 0 corresponds to an arbitrary time instant before the hovering and subsequent trampolining initiates. The corresponding FFT showing the dominant frequencies of droplet oscillation (marked by red arrows) in (b) regime 1: the decaying bouncing regime, and (c) regime 3: reemergence of droplet trampolining and bouncing.*

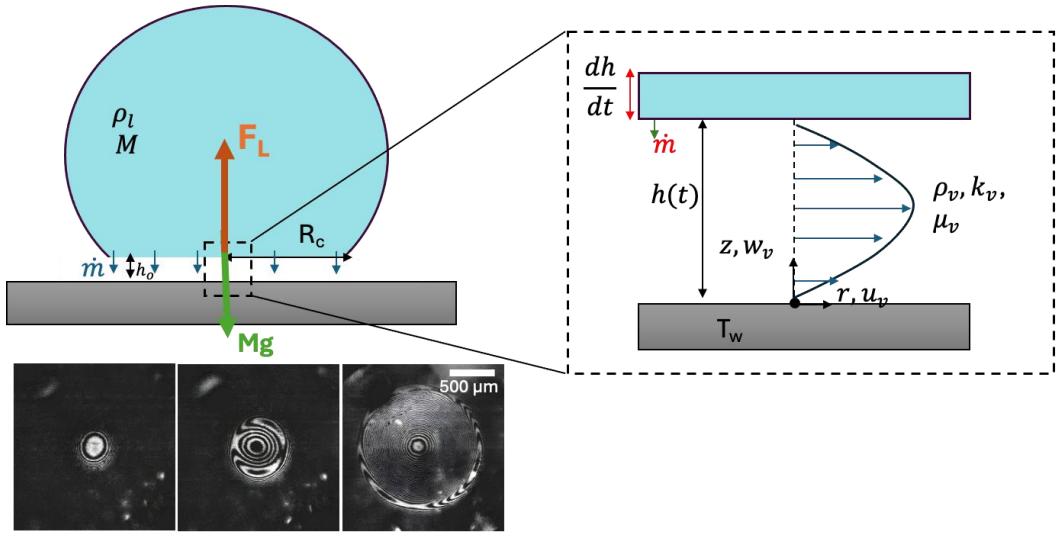

*Figure 4: Force balance for a LF droplet levitating on the vapor layer, (inset) control volume analysis for flow in the vapor layer; (bottom) interferometric images as the droplet contacts the heated substrate.*

For a static LF droplet under equilibrium, the weight of the LF droplet is balanced by the lubrication force due to flow in the vapor layer. For a small displacement of the LF drop below the equilibrium height, the decrease in the vapor layer thickness is accompanied by an increase in vaporization due to the reduction in thermal resistance. An increase in the vaporization is expected to result in an increase in the vapor flow force that pushes the droplet upwards that then comes down due to gravity. This cycle continues till the droplet completely evaporates.

The interplay between the dynamics of the intervening vapor layer and the droplet is expected to determine the underlying physics of droplet trampolining. We assume the droplet to be a levitating body of mass $M = \rho_l \frac{4}{3}\pi R^3$ at temperature $T_{sat}$, with an evaporation mass flow rate, $\dot{m}$ and a flat interface at the base[6]; here $R$ is the equivalent radius of the droplet. At any time instant, $t$, evaporation primarily occurs at the bottom surface of the droplet over a contact radius, $R_c$ due to heat conduction through the vapor layer of thickness $h(t)$ (Figure 4 (inset)). For a temperature difference, $\Delta T = T_w - T_{sat}$ across the vapor layer, the evaporation mass flow rate can be written as, $\dot{m} = \dot{m}''A_c = \frac{k_v \Delta T}{h_{fg} h} A_c$ where $A_c = \pi R_c^2$, $k_v$ is the vapor thermal conductivity, and $h_{fg}$ is the latent heat of vaporization. The contact radius can be approximated as the Hertz contact length, $R_c = \frac{R^2}{l_c}$ [3,20].

Performing a control volume analysis for the droplet, the linear momentum conservation in the vertical direction is given as

$$\frac{d}{dt}(Mv) + \dot{m}(v - v_v) = F_L - Mg \qquad (1)$$

where, $M$ is the instantaneous mass of the droplet, $v = \frac{dh}{dt}$ is the velocity of the droplet, $v_v$ is the vapor exit velocity at the base of the droplet, $F_L$ is the lubrication force due to the vapor film, and $g$ is the acceleration due to gravity. Displacement of the droplet, $h$, is taken as the height of the liquid-vapor interface from the substrate. The force, $F_L$, is calculated from the Navier-Stokes (NS) equations by using lubrication approximation since $\left(\frac{h}{R_C}\right)^2 Re \ll 1$ where $Re$ is the Reynolds number of the vapor flow as (see SI.S7),

$$F_L = \frac{3\mu \pi R^4}{2h^3}\left(-\frac{dh}{dt} + \frac{\dot{m}''}{\rho_l}\right) \qquad (2)$$

The governing equation for the height ($h$) of the liquid-vapor interface from the substrate is obtained from eq 1 as,

$$\frac{d^2h}{dt^2} + \left(\frac{1}{M}\frac{3\pi}{2}\frac{\mu_v R_c^4}{h^3}\right)\frac{dh}{dt} + \left(g - \frac{\dot{m}''^2 \pi R_c^2}{\rho_v M} - \frac{3\pi \mu_v R_c^4 \dot{m}''}{2\rho_v h^3 M}\right) = 0 \,. \qquad (3)$$

The above equation is analogous to a nonlinear spring-mass-damper system. The equilibrium height, $h_o = \left(\frac{9}{8}\frac{\mu_v k_v \Delta T}{h_{fg} \rho_v}\frac{\rho_l g}{\gamma^2}\right)^{\frac{1}{4}} R^{\frac{5}{4}}$ is the stationary solution of the above nonlinear ordinary differential equation and defines the vapor gap of the LF droplet at rest.

Nondimensionalizing eq 3 using $h_o$ and $\tau = \sqrt{\frac{h_o}{g}}$ as the characteristic length and time scales, respectively, and defining $\tilde{h} = h/h_o$ and $\tilde{t} = t/\tau$, we get,

$$\frac{d^2\tilde{h}}{d\tilde{t}^2} + \frac{D}{\tilde{h}^3}\frac{d\tilde{h}}{d\tilde{t}} + \left(1 - \frac{\tilde{F}_v}{\tilde{h}^2} - \frac{\tilde{F}_l}{\tilde{h}^4}\right) = 0 \quad (4)$$

Here, $\tilde{F}_v = \frac{3R_c^2}{4h_o^2}\left(\frac{k_v \Delta T}{h_{fg}}\right)^2\left(\frac{1}{R^3 \rho_l \rho_v g}\right) = \frac{\dot{m}''^2 A_c/\rho_v}{Mg}$ is the ratio of the vapor recoil force and weight of the droplet, $\tilde{F}_l = \frac{9}{8}\frac{\mu_v R_c^4}{R^3 \rho_l \rho_v g h_o^4}\frac{k_v \Delta T}{h_{fg}} = \frac{3\pi}{2}\frac{\mu_v R_c^4 \dot{m}''}{h_o^3 \rho_v}\left(\frac{1}{Mg}\right)$ is the ratio of the lubrication lift and weight of the droplet, and $D = \frac{9\mu_v R_c^4}{8h_o^3 R^3 \rho_l}\sqrt{\frac{h_o}{g}} = \frac{3\pi}{2}\frac{\mu_v R_c^4}{h_o^3 M}\sqrt{\frac{h_o}{g}}$ is the nondimensionalized lubrication damping coefficient.

Eq. 4, with constant parameters $D$, $\tilde{F}_v$ and $\tilde{F}_l$, is expected to result in bouncing with an amplitude smaller than the initial height of the liquid-vapor interface (see SI S7). We note that the contact radius, $R_c$, assumed to be a constant in the above derivation, is in fact subjected to temporal variations due to small oscillations of the droplet. The primary mode of the droplet oscillation is expected to be the second Legendre mode and the corresponding nondimensional natural frequency is given by $\tilde{f}_{Rayleigh} = f_{Rayleigh}\sqrt{\frac{h_o}{g}}$. Therefore, to account for these oscillations, we can assume $R_c = \overline{R_c}(1 + \varepsilon\cos(2\pi\tilde{f}_{Rayleigh}\tilde{t}))$, where $\overline{R_c}$ is the mean contact radius and $\varepsilon$ is a small parameter. The parameters $\tilde{F}_v$, $\tilde{F}_l$, and $D$ are proportional to $R_c^2, R_c^4$ and $R_c^4$, respectively. We substitute this time dependence of $R_c$ in eq 4 and linearize in $\varepsilon$, retaining the $O(\varepsilon)$ terms. Further, to investigate the stability of the vapor thickness, we perturb the mean vapor thickness as $\tilde{h} = 1 + \xi$, where $\xi \ll 1$ is a small perturbation. Neglecting the nonlinear terms in $\xi$, we get a linearized second order differential equation for $\xi$ as

$$\frac{d^2\xi}{d\tilde{t}^2} + D\big(1 - 4\varepsilon\cos(2\pi\tilde{f}_{Rayleigh}\tilde{t})\big)\frac{d\xi}{d\tilde{t}} + \Big((2\tilde{F}_v + 4\tilde{F}_l) + 16\tilde{F}_l\varepsilon\cos(2\pi\tilde{f}_{Rayleigh}\tilde{t})\Big)\xi$$

$$= (2\tilde{F}_v + 4\tilde{F}_l)\,\epsilon\,\cos(2\pi\tilde{f}_{Rayleigh}\tilde{t}) \quad (5)$$

Since the parameters $D \sim 10^{-1}$ and $\tilde{F}_v \sim 10^{-4}$ are significantly smaller compared to the typical values of the parameter $\tilde{F}_l \sim 1$ (for a typical droplet size of 1 mm), we ignore the damping and vapor thrust terms in the subsequent analysis. Thus, the above equation reduces to a non-homogeneous Mathieu's equation given by

$$\frac{d^2\xi}{d\tilde{t}^2} + (4 + 16\varepsilon \cos(2\pi \tilde{f}_{Rayleigh}\tilde{t}))\xi = 4\epsilon \cos(2\pi \tilde{f}_{Rayleigh}\tilde{t}) \quad (6)$$

For $\epsilon = 0$, natural frequency of vibration of the vapor film is 2, corresponding to dimensional frequency $\frac{1}{\pi}\sqrt{\frac{g}{h_o}}$ (see SI.S7).

Thus, nondimensional frequencies at which a solution of eq. 6 is unstable to infinitesimal perturbations (*i.e.,* tongues of Mathieu's equation)[25], are

$$\frac{4}{(2\pi \tilde{f}_{Rayleigh})^2} = N^2, \quad (7)$$

where, $N = 0, ½, 1, 3/2, 2, 5/2,....$

In dimensional form, we get:

$$N = \frac{f_v}{f_{Rayleigh}} = \frac{1}{\pi}\left(\frac{g}{h_o}\right)^{\frac{1}{2}}\left(2.65\sqrt{\frac{\rho R^3}{\gamma}}\right) \quad (8)$$

where, the natural frequency of the vapor film thickness is given by $f_v = \frac{1}{\pi}\left(\frac{g}{h_o}\right)^{\frac{1}{2}}$. For $N = 0, 1, 2, \ldots$ the response of the vapor layer thickness is harmonic whereas, for $N = 1/2, 3/2, 5/2,\ldots$, the response is subharmonic at $f_{Rayleigh}$. At these specific values of $N$, the liquid-vapor interface is unstable and the droplet starts bouncing with increased amplitude (see SI S8).

The ratio $N$ depends on the liquid and vapor properties at the saturation temperature, the substrate superheat, and the size of the droplet. The natural frequency of a water droplet ($f_{Rayleigh}$) and the intervening vapor layer ($f_v$), and $N$ (from eq 8) as a function of droplet radius are shown in Fig 5. The experimentally observed radii (marked by star symbols in Fig 5) for the occurrence of trampolining correspond to the integer and half integer values of $N$. For instance, the radius at which $f_v = f_{Rayleigh}$ (i.e. $N = 1$) is in agreement with the experimental observation of droplet size ($R = 0.33$ mm) at which trampolining is observed. Therefore, we infer that the observation of trampolining events at different sizes of the evaporating droplet (Fig 5) is due to the parametric resonance caused by the subharmonic/harmonic coupling between the droplet oscillations and the intervening vapor layer.

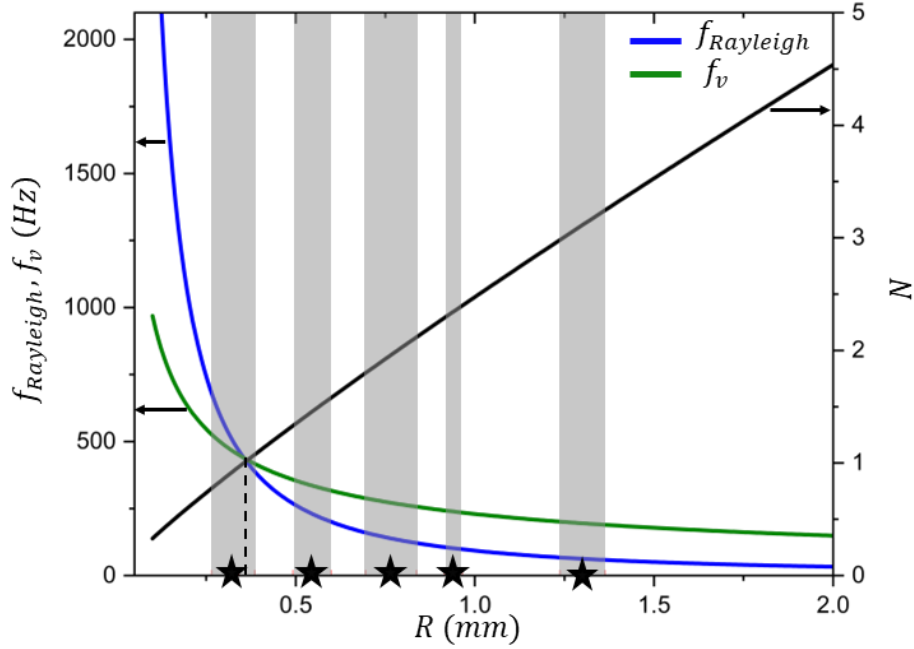

*Figure 5: $f_{Rayleigh}$, $f_v$ and $f_v/f_{Rayleigh}$ as a function of droplet radius; the marked star symbols and the shaded region on the radius axis correspond to the experimental average radii and corresponding standard deviation at which trampolining is observed.*

The theoretical value of the droplet radius ($R_{th}$) at which the resonance-driven trampolining occurs is,

$$R_{th} = \left(\frac{N\pi}{2.65}\right)^{8/7} \left(\frac{9\mu_v k_v \Delta T \, l_c^4}{8 h_{fg} \rho_v \rho_l g}\right)^{\frac{1}{7}} = \left(\frac{N\pi}{2.65}\right)^{8/7} \left(\frac{h_{0,l_c}}{l_c}\right)^{\frac{4}{7}} l_c , \qquad (9)$$

where, $h_{0,l_c}$ is the equilibrium vapor thickness evaluated for a droplet of radius $l_c$.

The substrate superheat has a weak influence on the droplet radius at which trampolining is observed since $R_{th} \sim \Delta T^{\frac{1}{7}}$ (eq 9; SI Fig S6). The ratio of the theoretical and experimental trampolining radius, $R_{exp}/R_{th}$ is plotted for different frequency ratio $N$ to provide a comparison between the experimental data and the theoretical predictions (Fig 6). We also plot the radii of the water droplet at which increase in the bouncing amplitude was observed by Liu and Tran[22] in Fig 6 and show the agreement of their experimental observation with our theory. The finding $R_{exp}/R_{th} \sim 1$ irrespective of the liquid properties (volatility, viscosity, surface tension, density and other the thermophysical properties) suggests a universality of the parametric resonance induced trampolining behavior of the levitating droplets.

In summary, as the droplet evaporates and scans different radii, the droplet undergoes subharmonic and harmonic coupling between the vapor flow underneath and the droplet oscillations resulting in parametric resonance that causes them to undergo trampolining intermittently for specific radii. Parametric resonance in this system arises from temporal variations in the contact area of the levitating droplet leading to time dependent lubrication force. Remarkably, this mechanism demonstrates that trampolining can occur even at small droplet radius ($\sim 0.1 l_c$). The mechanism explains the dynamics of bouncing and trampolining for the entire range of radii of LF droplets of different liquids without considering the effect of internal flow and the exact shape of the liquid-vapor interface.

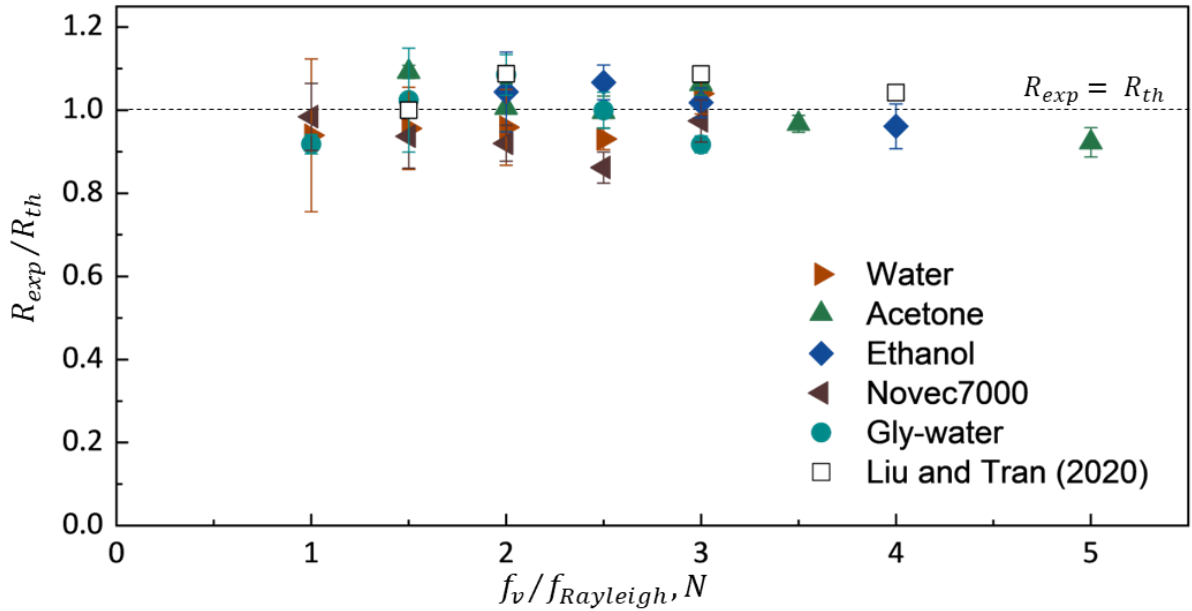

*Figure 6: Comparison of the experimental and theoretical values of the multiple reemergence radius for different liquids.*

**Materials and Methods**

The experiments are conducted on a transparent quartz, and steel concave substrate with radius of curvature ~50 mm and ~26 mm respectively. The slight curvature is provided to ensure that the droplet stays confined despite its high mobility. The experiments are performed at temperatures above the Leidenfrost (LF) temperature of the liquid on the substrate. The test substrate is heated using a PID-controlled heater (WATLOW) and the temperature of the surface is measured through a K-type leaf thermocouple (Delta Power). The temperature of the substrate is varied from 200-400 °C depending on the LFT of the liquid used.

Different liquids such as DI water, water-glycerol mixtures (10 wt % Gly (gly10), 25 wt % Gly (gly25), 50 wt% Gly (gly50)), acetone, ethanol, and Novec-7000 are used for the experiments. A droplet of a given volume is gently dispensed on the superheated substrate using a calibrated micropipette (Thermoscientific). In addition to the investigation of the overall droplet bouncing dynamics, flow visualization inside the droplet is conducted using $d_p = 10$–$13$ $\mu m$ hollow glass spheres (Sigma-Aldrich) as seeding particles. Considering the particles to be spherical, the settling velocity given by Stokes law is expressed as, $v_p = (\rho_p - \rho)gd_p^2/18\mu = 0.3 \times 10^{-3}$ $mm/s$ where ($\rho_p$= 1050 kg/m³) is the density of particles and $\rho = 958$ $kg/m^3$ and $\mu = 0.5$ mPa-s are the density and dynamic viscosity of water. This velocity is significantly less compared to the experimentally obtained fluid velocity inside the hovering droplet, which is the range of $5 - 50 \times 10^4$ $mm/s$. The internal flow visualization is done through a bulk front side illumination of the droplet using optic guide light source (Dolan Jenner MI 152) to obtain qualitative information regarding the flow inside the droplet. The evaporation and the bouncing behavior of the LF droplet is visualized using a high-speed camera (Photron Fastcam mini AX200) at 125 -1000 fps via shadowgraphy with backlighting (Zaila LED light). All the experiments are repeated at least 3 - 5 times to ensure repeatability of the observations and calculating the average radius at which trampolining is observed. For high-speed videos (1000 fps), an end trigger is used such that at least a single trampolining event is captured. The high-speed videos are analyzed using an in-house MATLAB code to determine the transient variation in droplet size and determine the bouncing heights and frequency.

**High speed interferometry experiments**

For the interferometric visualizations, simultaneous side view and bottom view high speed imaging are performed. A pulsed diode laser (Cavitar CAVILUX laser, 632 nm) is used for interferometry and a LED light source (Zaila) for backlighting are triggered simultaneously during the experiments. A 4x objective lens (Olympus) is used along with the Navitar lens providing a resolution of 0.44 micron/pixel; the fringe width varies from ~ 3 - 12 pixels depending on the size of the droplet. The vapor layer characterization is done by the fringe counting method after the initial image processing in MATLAB.

**Acknowledgement**

We gratefully acknowledge the insightful discussions with Prof. David Quéré.